# The Effect of TCP Variants on the Coexistence of MMORPG and Best-Effort Traffic

Jose Saldana[1], Mirko Suznjevic[2], Luis Sequeira[1], Julián Fernández-Navajas[1], Maja Matijasevic[2], José Ruiz-Mas[1]
[1]Communication Technologies Group (GTC) – Aragon Institute of Engineering Research (I3A)
Dpt. IEC. Ada Byron Building. CPS Univ. Zaragoza. 50018 Zaragoza, Spain
e-mail: {jsaldana, sequeira, navajas, jruiz}@unizar.es

[2]University of Zagreb Faculty of Electrical Engineering and Computing
Unska 3, 10000 Zagreb, Croatia
e-mail: {mirko.suznjevic, maja.matijasevic}@fer.hr

*Abstract-* We study TCP flows coexistence between Massive Multiplayer Online Role Playing Games (MMORPGs) and other TCP applications, by taking World of Warcraft (*WoW*) and a file transfer application based on File Transfer Protocol (FTP) as an example. Our focus is on the effects of the sender buffer size and FTP cross-traffic on the queuing delay experienced by the (MMORPG) game traffic. A network scenario corresponding to a real life situation in an ADSL access network has been simulated by using NS2. Three TCP variants, namely TCP SACK, TCP *New Reno*, and TCP *Vegas*, have been considered for cross-traffic. The results show that TCP *Vegas* is able to maintain a constant rate while competing with the game traffic, since it prevents packet loss and high queuing delays by not increasing the sender window size. TCP SACK and TCP *New Reno*, on the other hand, tend to continuously increase the sender window size, thus potentially allowing higher packet loss and causing undesired delays for the game traffic. In terms of buffer size, we have established that smaller buffers are better for MMORPG applications, while larger buffers contribute to a higher overall delay.

I. INTRODUCTION

Massive Multiplayer Online Role Playing Games (MMORPGs) are clearly (soft) real-time interactive applications. The fact that they typically use Transmission Control Protocol (TCP) as a transport protocol of choice is a somewhat surprising fact, which makes it interesting to study the behavior of MMORPG traffic when sharing a common link with other TCP applications.

True to their name, MMORPGs are all about role-playing, i.e., about developing one's virtual character (also known as *avatar*) over a long-term (virtual) life in a persistent virtual world, through various activities and acquiring knowledge, abilities, gadgets, weapons, etc. In terms of popularity, the leading MMORPG is the *World of Warcraft* (*WoW*) by Activision Blizzard, which, according to *mmodata.net*, has over 10 million players worldwide. While the level of players' activity and mutual interactions within a typical MMORPG is not as high as in First Person Shooter games (FPSs), which typically use UDP [1], MMORPGs are still considered a "real-time" service, with rather tight requirements in terms of timeliness of packet delivery. In comparison with FPSs, at the game traffic level there are many differences as well, for example, user behavior, session characteristics and dynamics, and the need for a reliable transport protocol to maintain the state of the game. As MMORPGs use TCP, game designers rely on its ability to reliably deliver the packets from one network host to another, over the network. The key issue from the users' perspective, however, is not (just) the end-to-end throughput, as much as the end-to-end delay ("lag"), due to its negative effect onto players' in-game performance and, thus, the user-perceived quality [2][3].

Being one of the fundamental Internet protocols, TCP is a rather complex, stateful, closed control loop protocol. Its behavior and performance have been studied extensively. Over the years, a number of TCP *variants* have been introduced to improve it. (For the purposes of this paper, the reader is assumed to have some familiarity with TCP variants and respective control variables, flow control, retransmission, and congestion control mechanisms.) In this work, we focus on a scenario related to user perceived quality. By using NS2 simulation, we study the queuing delay experienced by the MMORPG game traffic as a result of flows competing for link resources with "other" present TCP flows (here, FTP) in the network. We compare the protocol behavior when using different TCP variants, namely, "standard" TCP SACK, TCP *New Reno*, and TCP *Vegas*. We also examine the effects of the sender socket buffer.

The paper is organized as follows: Section II discusses related work, and Section III covers the simulation scenarios. The results are presented in Section IV. Section V concludes the paper.

II. RELATED WORK

*A. Modelling the traffic of MMORPGs*

In this section we briefly survey related research efforts in the areas of traffic modelling for games. Most of the papers follow the approach introduced by Paxon in [4], and initially used in the area of multiplayer games by Borella [5], who analysed the traffic of the FPS game *Quake*. Using this approach, Svoboda *et al.* [6], modelled the traffic of *WoW* in NS2, based on a trace obtained within a 3G mobile core network. Wu *et al*. modelled the traffic of one of the most popular MMORPGs in China: *World of Legend* [7]. Their model is based on a trace obtained on a client accessing the game through a mobile GPRS access network. Kim *et al.* [8] presented the traffic models of the MMORPG *Lineage,* based on a trace obtained from the game provider.

Several authors use this approach, combined with the classification of application level behaviour, in order to better model the highly erratic traffic of MMORPGs. Wang *et al.* [9], proposed the following classification: *Downtown*, *Hunting*, and *Battlefield*, and presented traffic models for NS2. They also investigated characteristics of traffic in different access networks. Park *et al.* proposed the following classification of *WoW* situations: *Hunting the NPCs*, *Battle with players*, *Moving*, and *No play* [10] and modelled the traffic across each class. Suznjevic *et al.* [11] classified the states of the virtual world through categories of user actions focused on the player progression, namely: *Questing, Trading, Player vs Player combat, Dungeons,* and *Raiding*. A network traffic model for each category was also provided. Finally, a study of the subjective quality which can be achieved for an MMORPG, depending on network parameters, was conducted in [12], where a MOS model was presented for *WoW*.

### B. TCP variants

TCP variants handle congestion avoidance in different ways. TCP *New Reno* and TCP SACK use packet loss information to reduce their sending rate. TCP *Vegas* uses the RTT increase so as to prevent packet loss [13]. However, an analysis of TCP *Vegas* [14] showed that it may be unfair and may also increase the number of queued packets if network buffering capacity is small. TCP *hybla* [15] is able to achieve RTT fairness between a number of flows, in certain conditions. In [16], TCP *Libra* was proposed, with the aim of avoiding the problem of RTT unfairness, introducing *scalability* and *penalty* factors.

The problem of the coexistence of real-time and non real-time flows using TCP and UDP in the same network has also been studied. In [17] an anomalous zone in which packet loss of UDP flows grew with the increase of buffer size was observed when sharing the buffer with TCP connections; in [18] the coexistence of real-time UDP flows and different TCP variants was studied in a home network, proposing an enhanced access point. The use of TCP for video delivery was also evaluated in [19], where TCP *Westwood* was proposed as suitable for a wireless home scenario.

A study of the effect of different TCP versions on *Anarchy Online* MMORPG was performed in [20], showing that no significant improvement is possible when changing the TCP variant of the game traffic. The study presented in this paper differs from previous works, as its main aim is to analyze the effects of FTP cross-traffic using different TCP variants on the TCP traffic of an MMORPG.

## III. SIMULATION METHODOLOGY

### A. Network scenario

For the purposes of our study, we consider a scenario with two home users (using a wired LAN in the same household) sharing the ADSL access link while performing different activities at the same time: one user is playing a MMORPG, and the other is running a large file upload. We use NS2 (http://www.isi.edu/nsnam/ns/) as a simulation tool. A dumbbell topology, shown in Fig. 1, in which a TCP game flow shares a common link with a TCP flow from the FTP application is used. Both the gaming and the FTP sessions have been established from the respective user, through the local network, over the common link, to a corresponding server located in the Internet, and run simultaneously.

For the simulated home network, the network parameters have been selected to fit a residential asymmetric (ADSL) access network, with 512 kbps uplink, and 6 Mbps downlink. The speed of local network is 100 Mbps. Two buffer sizes have been selected to be simulated for the uplink: a) 200 packets, and b) 20 packets. The total end to end one-way delay of the simulated network has been set to 80 ms, considering an inter-region scenario [22]. For the synthetic game traffic generation, we use World of Warcraft (*WoW*), generated according to the known statistical models [6], [21]. *WoW* traffic uses TCP SACK variant. For the FTP background application, we use the NS2 built-in traffic generation model, and three different TCP variants, namely TCP SACK, TCP *New Reno*, and TCP *Vegas*.

### B. Characteristics of the MMORPG traffic

MMORPG traffic is simulated through two flows, one carrying the actions of the player (client) to the server, and the other communicating the current state of the game from the server to the client application. The MMORPG client/server communication does not require a lot of bandwidth; it has been shown that, depending on the situation in the virtual world, traffic demands may rise up to 50 kbps in downlink and 5 kbps in uplink [21], but that the average demands are typically lower [6].

Several traffic models have been proposed in the literature for inter-packet time and Application Protocol Data Unit (APDU) sizes. In this work, we use the model proposed in [11], and the most frequent category, i.e., *Questing*. (The histograms for inter-packet time and APDU can be found in [11] as well). The game generates small packets, since the majority of the APDUs are small and, in addition, the PUSH bit of the TCP header is set to 1, in order to make the protocol send it as soon as possible (as opposed to waiting for a larger amount of data to be transmitted).

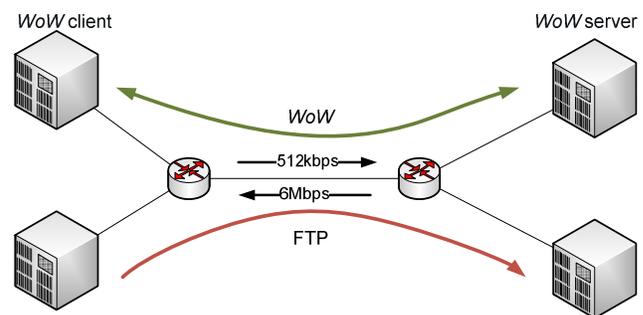

Figure 1. NS2 simulation topology and scenario.

## C. Generation of the traffic for the simulation

First, an NS2 script has been created, which generates APDUs with the corresponding statistical distribution for *Questing* activity in *WoW*, according to the model presented in [11] (eight different values for the client, and a Lognormal distribution for the server), and also with the correspondent inter-arrival times for the APDUs (two Weibull distributed and two deterministic values for the client, and one Normal, one Weibull and three deterministic values for the server).

The simulated *WoW* application uses the TCP stack in the client (player) computer. For the simulation purposes, we have selected TCP SACK, as it is a TCP version running in many home computers [19]. The simulator calculates the inter-packet time and the APDU size and, if necessary, it divides the APDU into 1,460 bytes chunks, which are sent in 1,500 bytes packets. ACKs are also sent by the TCP stack, according to the SACK implementation of NS2. Fig. 2 shows the Cumulative Distribution Functions (CDFs) for *WoW* packet sizes and inter-packet times, once the traffic has been sent. It can be observed that many 40-byte packets are present, corresponding to the ACKs. Inter-packet time is small, roughly 155 ms on average.

For FTP we used NS2 built-in application, with 1,500 byte packets (1,460 byte payload) and three TCP variants, as noted earlier – TCP SACK, TCP *New Reno*, and TCP *Vegas*.

The simulation time has been set to 1,000 seconds.

## IV. SIMULATION RESULTS

Figs. 3 to 5 show the TCP window size for (a) *WoW* and (b) FTP, for the 200-packet buffer. Figs. 6 to 8 show the respective values for the 20-packet buffer. The results for the uplink queuing delay for *WoW* are shown in Fig. 9. The average over the last 200 seconds of the simulation is used.

Comparing the figures, it is clear that the sending window of *WoW* behaves differently from that of FTP. The reason for this is that the game application is not trying to consume as much available bandwidth as possible (as FTP does). It only has to send a continuous data flow of small packets, at a rate of less than 10 kbps. Hence, the window size varies, depending on the burstiness of the data generated by the game application, the maximum size being about 20 packets.

First, we discuss the 200-packet buffer scenario. A 200-packet buffer can be considered relatively "large" when dealing with real-time applications – in [23], using "tiny" buffers of few tens of kilobytes was recommended to keep the delay at minimum. Figs. 3 and 4 show that the window of FTP continuously grows when TCP SACK and TCP *New Reno* are used. While a larger buffer on the uplink has a positive effect of preventing packet loss occurrence, on the negative side it also adds delays that are unacceptable from the players' perspective [12]. As shown in Fig. 9: the added delays using TCP SACK and TCP *New Reno* is nearly 300 ms.

When using TCP *Vegas* for the background flow (Fig. 5), however, the window size remains stable. As reported in [16], the cause of this "timid" behavior of TCP *Vegas* is that it notices the RTT increase, so it maintains its sending rate at the same level, even when there is no packet loss.

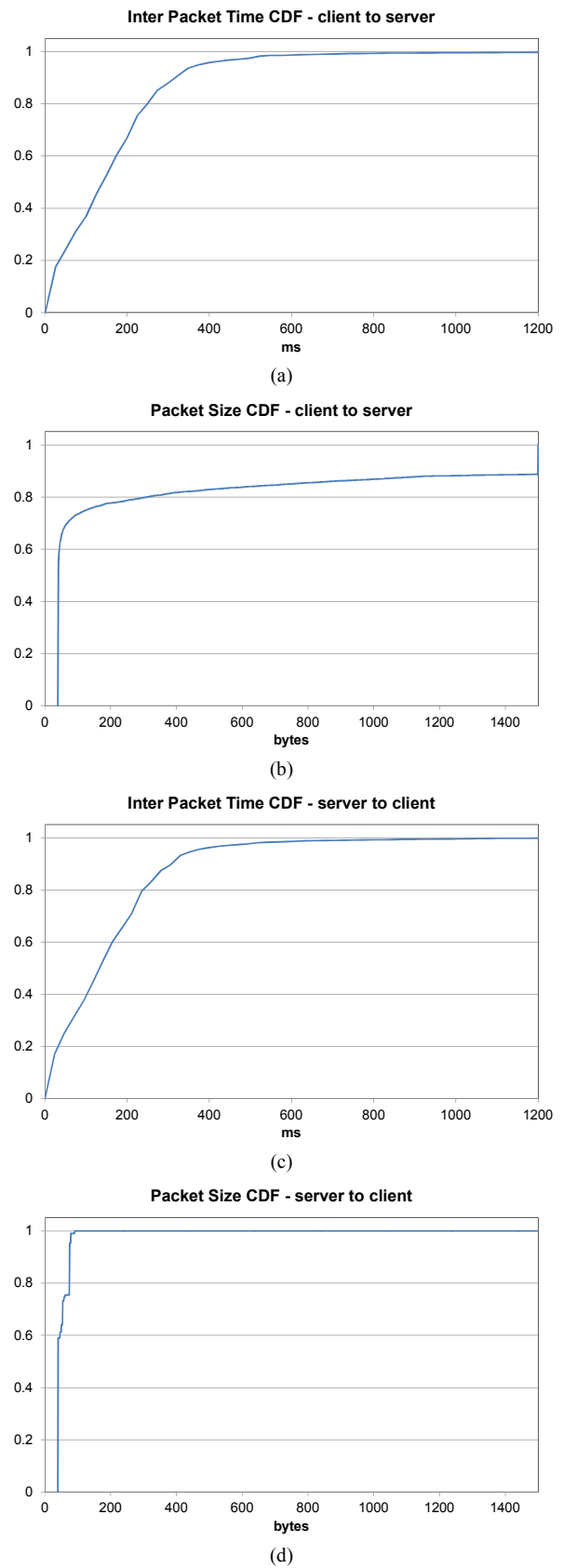

Figure 2. Inter Packet Time and Packet size CDFs for client-to-server and server-to-client traffic (using TCP SACK).

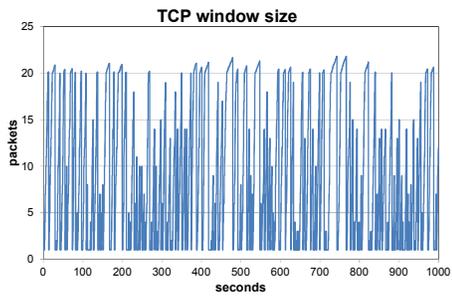
(a)

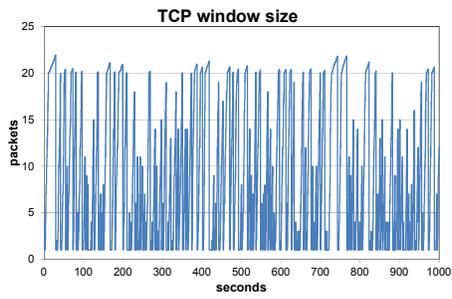
(a)

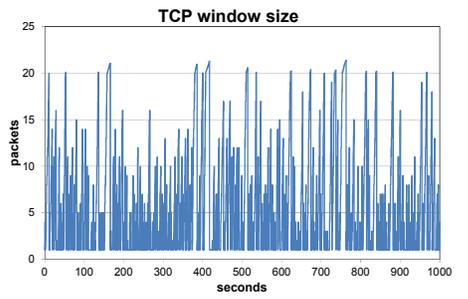
(a)

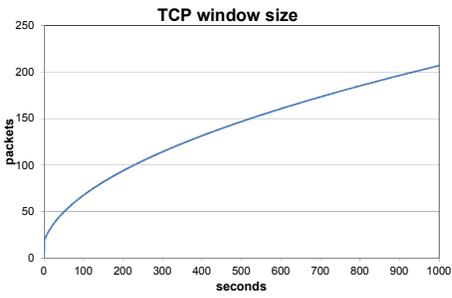
(b)

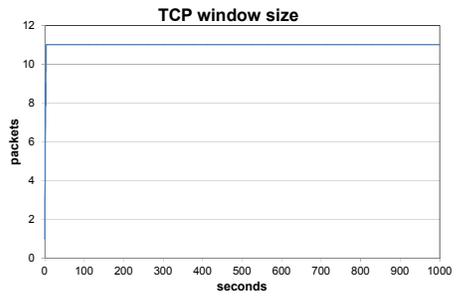
(b)

Figure 3. TCP window size with 200 packets buffer, when using TCP SACK for the background FTP source: a) *WoW*; b) background traffic

Figure 4. TCP window size with 200 packets buffer, when using TCP *New Reno* for the background FTP source: a) *WoW*; b) background traffic

Figure 5. TCP window size with 200 packets buffer, when using TCP *Vegas* for the background FTP source: a) *WoW*; b) background traffic

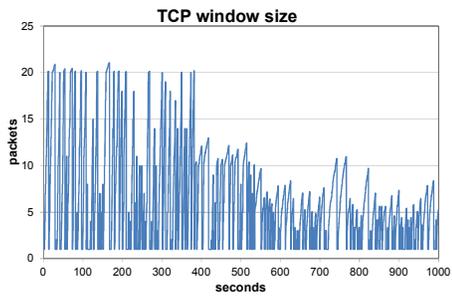
(a)

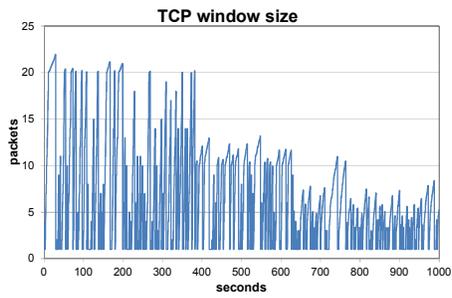
(a)

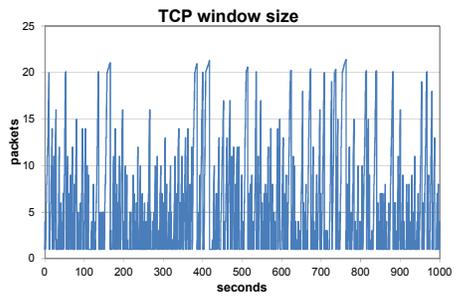
(a)

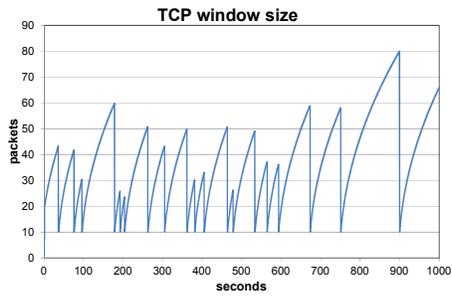
(b)

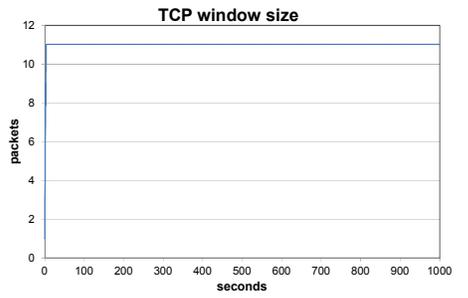
(b)

Figure 6. TCP window size with 20 packets buffer, when using TCP SACK for the background FTP source: a) *WoW*; b) background traffic

Figure 7. TCP window size with 20 packets buffer, when using TCP *New Reno* for the background FTP source: a) *WoW*; b) background traffic

Figure 8. TCP window size with 20 packets buffer, when using TCP *Vegas* for the background FTP source: a) *WoW*; b) background traffic

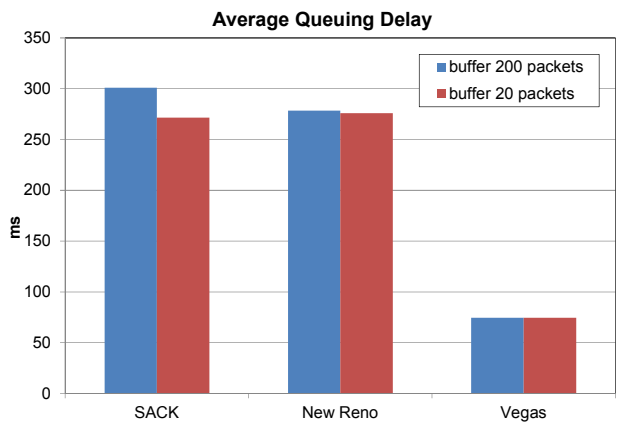

Figure 9. Queuing delay of *WoW* traffic (always using TCP SACK) when using different TCP versions for the background FTP source

For the smaller, 20-packet buffer, there is some packet loss, so both TCP SACK and TCP *New Reno* adapt their sending window accordingly. Interestingly, as they also keep on increasing their rate, they affect the *WoW* flow, which reduces its window (Figs. 6 and 7). TCP *Vegas* (Fig. 8), on the other hand, has a similar behavior to that for the larger, 200-packet buffer: it maintains its window size, which results in smaller delays for the game traffic, as shown in Fig. 9.

## V. CONCLUSIONS

The results show that TCP *Vegas* is able to maintain a constant rate while competing with the game traffic, since it prevents packet loss by avoiding the increase of the sending window size. Opposite to that, TCP SACK and TCP *New Reno* tend to keep on increasing the window size, thus adding undesired delays to the game traffic. Finally, smaller buffers have been demonstrated to be better for TCP-based MMORPGs, since larger buffers cause higher delays.

As future work, we plan to run simulations with more recent TCP variants (e.g., TCP *Libra*), both for gaming and for background traffic, in order to check its suitability in these scenarios, also taking into account their coexistence with widely deployed TCP variants.

## ACKNOWLEDGMENT

This work has been partially financed by CPUFLIPI Project (MICINN TIN2010-17298), the European Community's 7th Framework Programme under grant agreement no. 285939 (ACROSS), MBACToIP Project, of Aragon I+D and Ibercaja Obra Social, Project Catedra Telefonica, Univ. Zaragoza.